\RequirePackage{fix-cm}
\documentclass{article}

\usepackage{PRIMEarxiv}
\usepackage[utf8]{inputenc} 
\usepackage[T1]{fontenc}    
\usepackage{hyperref}       
\usepackage{url}            
\usepackage{booktabs}       
\usepackage{amsfonts}       
\usepackage{nicefrac}       
\usepackage{microtype}      
\usepackage{lipsum}
\usepackage{fancyhdr}       
\usepackage{graphicx}       
\graphicspath{{media/}}     

\usepackage{tikz}
\usepackage{amsmath}
\usepackage{xfrac}

\usepackage{adjustbox}
\usepackage{soul}
\usepackage{subcaption}

\usepackage{listings}
\usepackage{xcolor}       
\usepackage{multirow}
\usepackage[numbers]{natbib}

\usepackage{times}
\usepackage{latexsym}
\usepackage[most]{tcolorbox}
\usepackage{enumitem}

\colorlet{lightyellow}{yellow!40}
\definecolor{navyblue}{rgb}{0.0, 0.0, 0.5}

\makeatletter
{\small 
\xdef\f@size@small{\f@size}
\xdef\f@baselineskip@small{\f@baselineskip}
\normalsize 
\xdef\f@size@normalsize{\f@size}
\xdef\f@baselineskip@normalsize{\f@baselineskip}
}
\newcommand{\smalltonormalsize}{%
  \fontsize
    {\fpeval{(\f@size@small+\f@size@normalsize)/2}}
    {\fpeval{(\f@baselineskip@small+\f@baselineskip@normalsize)/2}}%
  \selectfont
}
\makeatother

\definecolor{forestgreen}{rgb}{0.13, 0.55, 0.13}

\usepackage{anyfontsize}

\title{Atomic Learning Objectives Labeling: \\A High-Resolution Approach for Physics Education}

\author{
  Naiming Liu \\
  Rice University \\
  \texttt{nl35@rice.edu} \\
  \And
  Shashank Sonkar \\
  Rice University \\
  \texttt{ss164@rice.edu} \\
  \AND
  Debshila Basu Mallick \\
  OpenStax \\
  \texttt{db19@rice.edu} \\
  \And
  Richard Baraniuk \\
  Rice University \\
  \texttt{richb@rice.edu} \\
  \And
  Zhongzhou Chen \\
  University of Centeral Flordia \\
  \texttt{zhongzhou.chen@ucf.edu} \\
}

\begin{document}
\maketitle

\begin{abstract}
This paper introduces a novel approach to create a high-resolution ``map" for physics learning: an "atomic" learning objectives (LOs) system designed to capture detailed cognitive processes and concepts required for problem solving in a college-level introductory physics course. 
Our method leverages Large Language Models (LLMs) for automated labeling of physics questions and introduces a comprehensive set of metrics to evaluate the quality of the labeling outcomes.
The atomic LO system, covering nine chapters of an introductory physics course, uses a ``subject-verb-object'' structure to represent specific cognitive processes. We apply this system to 131 questions from expert-curated question banks and the OpenStax University Physics textbook. Each question is labeled with 1-8 atomic LOs across three chapters.
Through extensive experiments using various prompting strategies and LLMs, we compare automated LOs labeling results against human expert labeling. Our analysis reveals both the strengths and limitations of LLMs, providing insight into LLMs reasoning processes for labeling LOs and identifying areas for improvement in LOs system design.
Our work contributes to the field of learning analytics by proposing a more granular approach to mapping learning objectives with questions. Our findings have significant implications for the development of intelligent tutoring systems and personalized learning pathways in STEM education, paving the way for more effective ``learning GPS'' systems.
\end{abstract}


\section{Introduction}
The rapid advancement of Artificial Intelligence (AI) has opened new avenues for developing intelligent learning systems that improve student learning outcomes \cite{sonkar2020qdkt, sonkar2023class, sonkar2024code}. An ideal learning system should function as a ``learning GPS,'' capable of assessing a student's current learning states and level of mastery, then guiding them towards the most suitable learning resources. 

However, current versions of a ``learning GPS'' suffers from ``low-resolution maps'' of knowledge domains, particularly in STEM disciplines. Existing learning objective systems, designed primarily for human labeling, often provide only broad descriptions of skills and concepts related to learning resources. For instance, in the OpenStax University Physics textbook, each sub-chapter is labeled with only 2-3 learning objectives, despite containing 20-40 end-of-chapter review questions. This low granularity limits the effectiveness of learning guidance, similar to providing vague driving directions. The fundamental reason behind the low-resolution map lies in the design of taxonomy systems designed for human domain experts. This approach not only restricts the number of objectives that can be tagged but also leads to potential degeneracy and ambiguity among learning objectives.

In this paper, we present a novel approach to create a higher-resolution ``map'' for physics learning: 1). We introduce an ``atomic'' learning objectives (LOs) system designed to capture detailed cognitive processes and concepts required for problem-solving. 2). We use Large Language Models (LLMs) for reliable automated labeling of learning resources. 3). We propose a comprehensive set of metrics to quantitatively and qualitatively evaluate the quality of LLMs labeling outcomes. Our atomic LO system, developed for a college-level introductory physics course, covers nine chapters. It builds upon previous work on physics taxonomy systems, such as NTEO~\cite{Teodorescu2013}. Each atomic LO describes a specific cognitive process in physics problem-solving using a constrained set of verbs and adheres to a ``subject-verb-object'' structure, representing ``input,'' ``cognitive action,'' and ``output.''

We apply the atomic LO system to a diverse set of 131 questions from expert-created problem banks for a physics course and OpenStax University-physics textbook. The questions are selected from three chapters and each question is tagged with 1-8 atomic LOs. We also conduct extensive experiments using various prompting strategies and LLMs to automate the LOs labeling process and compare the results against human expert labeling. Our analysis reveals both the strengths and limitations of LLMs in automated LOs labeling. We identify performance patterns that provide insights into LLM reasoning processes and highlight potential areas for improvement in both LLM capabilities and LO system design for AI applications. Our research contributes to the field of learning analytics by proposing a more granular approach to mapping learning objectives with questions, which paves the foundation for designing intelligent tutoring systems and personalized learning pathways in STEM education.

\section{Related Works}
\subsection{Learning Objectives Construction in Education}

Learning objectives are usually effective instructional design in education, with early approaches relying on manually developed taxonomies emphasizing on cognitive processes, such as Bloom's taxonomy~\cite{anderson2001taxonomy}. In physics education, curated systems often emphasize conceptual clarity and progression, such as the Force Concept Inventory~\cite{hestenes1992force}, which assesses student understanding of fundamental mechanics concepts. Other initiatives, like PhET simulations, integrate conceptual frameworks to enhance student engagement and comprehension~\cite{prima2018learning}.

\subsection{Automated Learning Objectives Labeling}

Expert labeling with learning objectives, while accurate, is impractical for large datasets due to variability in interpretation and potential fatigue. In contrast, automated methods for learning objective labeling and generation have emerged as useful tools that can facilitate human experts~\cite{balter2018estimating}. Previous research has explored mapping textbook materials and questions to concepts and skills using matrix factorization and VAE-based methods \cite{desmarais2013matrix, paaben2022sparse}, alongside studies focused on classifying knowledge components in mathematics and chemistry \cite{moore2024automated, tian2022automated}. Another line of research has applied topic modeling and hierarchical clustering to automatically derive learning objectives from course content \cite{gong2010comparing}. Recent advancements in natural language processing have opened new avenues for automating the process of identifying and labeling LOs. For instance, \cite{zur2023meta} uses GPT-3 to classify questions with LOs from OpenStax textbooks, while \cite{kwak2024bridging} explores the abilities of LLMs in tagging multilingual problem content with the appropriate skill from a taxonomy. LLMs have also shown promising results in various educational task, including automated grading, intelligent tutoring systems and student modeling \cite{liu2023novice, sonkar2023class, liu2022open, sonkar2024student}. However, LLM's application specifically for labeling or generating learning objectives in physics education remains a relatively unexplored area. Our research addresses this gap by utilizing LLMs to automate the creation and labeling of LOs for physics-related questions.

\section{Problem Formulation}
In this section, we provide the problem formulation for automated learning objective labeling using LLMs. Let $P = \{p_1, p_2, \ldots, p_n\}$ be a set of physics problems and $L = \{l_1, l_2, \ldots, l_m\}$ be a set of learning objectives. Define the ground truth mapping $G : P \rightarrow \mathcal{P}(L)$, where $\mathcal{P}(L)$ is a power set of $L$. Given a LLM $\mathcal{M}$, define a function $F_{\mathcal{M}} : A \times P \times \mathcal{P}(L) \rightarrow \mathcal{P}(L)$ such that

\[F_{\mathcal{M}}(a, p_{i}, S) = L' \subseteq S\]

where
\begin{itemize}
    \item $a \in A$ is a prompting strategy (e.g., simple, explanation, chain of thought)
    \item $p_{i} \in P$ is a physics problem
    \item $S \subseteq L$ is a subset of learning objectives provided to the LLM
    \item $L'$ is the subset of learning objectives predicted by the LLM for question $p_{i}$
\end{itemize}

For the learning objectives labeling task, we aim to

\begin{enumerate}
    \item Systematically assess the efficacy of various Large Language Models $\mathcal{M}$ and prompting strategies $a \in A$ in predicting learning objectives for physics problems. The assessment includes:
    \begin{itemize}
        \item Quantitative comparison of prediction accuracy across different model architectures and sizes
        \item Analysis of the impact of diverse prompting techniques on labeling performance
        \item Qualitative assessment of the models' ability to capture detailed learning objectives
        \item Identification of patterns, strengths, and limitations in the models' predictions
    \end{itemize}
    
    \item Formulate and develop comprehensive quantitative evaluation metrics $E : \mathcal{P}(L) \times \mathcal{P}(L) \rightarrow \mathbb{R}$ such that:
    \[E(F_{\mathcal{M}}(a, p_{i}, S), G(p_{i}))\]
    measures the agreement between the LLM predictions and the ground truth $G(p_{i})$ for question $p_{i}$. These metrics should: 
    \begin{itemize}
        \item Capture both exact matches and partial agreements in learning objective predictions
        \item Account for the hierarchical nature of learning objectives
    \end{itemize}
\end{enumerate}

\subsection{Evaluation Metrics}
We adopt four evaluation metrics to assess the performance of our learning objectives labeling system. For the definitions below, let $F$ be the set of predicted LOs and $G$ be the set of ground truth (human-labeled) LOs for a given question.

\subsubsection{Exact Match}
The Exact Match (EM) score is a binary metric that evaluates whether the predicted set of LOs exactly matches the ground truth set.
It is defined as:

\[
EM = \begin{cases}
1 & \text{if } F = G \\
0 & \text{otherwise}
\end{cases}
\]

\subsubsection{Jaccard Index}
The Jaccard Index (J), also known as the Jaccard similarity coefficient, quantifies the similarity between the predicted set of LOs and the ground truth set. It is defined as:

\[
J = \frac{|F \cap G|}{|F \cup G|}
\]

\subsubsection{F1 Score}
The F1 score balances precision and recall by calculating their harmonic mean:

\[
F1 = 2 \cdot \frac{\text{Precision} \cdot \text{Recall}}{\text{Precision} + \text{Recall}}
\]

where

\[
\text{Precision} = \frac{|F \cap G|}{|F|}, \quad \text{Recall} = \frac{|F \cap G|}{|G|}
\]

\subsubsection{Distance-based Metric}
\label{sec:eval_metrics}

We introduce a custom distance-based metric to capture the hierarchical nature of LOs and provide a more detailed evaluation of tagging accuracy. This metric considers the similarity between predicted and ground truth LOs at different levels of granularity. For each LO, we define:
\begin{itemize}
\item $LO^n$: LO Name (highest level, represents the general physics concepts)
\item $LO^a$: Action (represents the cognitive process of physics concepts)
\item $LO^c$: Code (represents the specific learning objective)
\end{itemize}

Examples of LOs with \textit{LO Name, Action and Code} can be found in Table~\ref{tab:los_example}.

We define a distance function $d(LO_1, LO_2)$ between two learning objectives as follows:

\[
d(LO_1, LO_2) = \begin{cases}
3 & \text{if } LO_1^n \neq LO_2^n \\
2 & \text{if } LO_1^n = LO_2^n \text{ and } LO_1^a \neq LO_2^a \\
1 & \text{if } LO_1^n = LO_2^n \text{ and } LO_1^a = LO_2^a \text{ and } LO_1^c \neq LO_2^c \\
0 & \text{if } LO_1^c = LO_2^c
\end{cases}
\]

For LOs that only appears in one set ($F$ or $G$), we define:

\[
d(LO, G) = \begin{cases}
1 & \text{if } LO^n \in \{LO'^n : LO' \in G\} \\
2 & \text{otherwise}
\end{cases}
\]

The overall distance-based score for a question is then calculated as:
\[
D = \sum_{LO \in F} \min_{LO' \in G} d(LO, LO') + \sum_{LO \in G \setminus F} d(LO, F)
\]
where $G \setminus F$ denotes the set difference.

\section{Dataset Construction}
Our dataset consists of two main components: LOs and questions. The dataset contains 9 chapters of LOs carefully curated to represent atomic-level cognitive processes in physics problem-solving, along with 131 questions selected for both expert and LLMs labeling. We also developed an interface that facilitates human experts to efficiently tag relevant LOs with each question.

\subsection{Structure of Learning Objectives}
\begin{table}[t!]
\begin{adjustbox}{width=0.95\textwidth, center}
\begin{tabular}{@{}p{0.95\textwidth}@{}}
\toprule
\vspace{-5mm}
\begin{minipage}[t]{\linewidth}
\begin{lstlisting}[mathescape=true,basicstyle=\ttfamily\normalsize]
$\textbf{LO Code}$:  $\colorbox{lightyellow}{\textbf{ME-KE-1}}$
$\textbf{LO Name}$:  Kinetic Energy (KE)
$\textbf{Item}$:     KE as scalar
$\textbf{Action}$:   Conc.ID
$\textbf{Provided}$: Situation or statments involving kinetic energy of a system
$\textbf{Outcome}$:  Correctly identify KE for an object, identify KE as non-direction, non-negative
\end{lstlisting}
\end{minipage} \\
\midrule
\vspace{-5mm}
\begin{minipage}[t]{\linewidth}
\begin{lstlisting}[mathescape=true,basicstyle=\ttfamily\normalsize]
$\textbf{LO Code}$:  $\colorbox{lightyellow}{\textbf{ME-KE-2}}$
$\textbf{LO Name}$:  Kinetic Energy (KE)
$\textbf{Item}$:     Magnitude of KE
$\textbf{Action}$:   Conc.Prop
$\textbf{Provided}$: Velocity of an object
$\textbf{Outcome}$:  Calculate magnitude of the kinetic energy 
\end{lstlisting}
\end{minipage} \\
\midrule
\vspace{-5mm}
\begin{minipage}[t]{\linewidth}
\begin{lstlisting}[mathescape=true,basicstyle=\ttfamily\normalsize]
$\textbf{LO Code}$:  $\colorbox{lightyellow}{\textbf{ME-GPE-2}}$
$\textbf{LO Name}$:  Gravitational PE (GPE)
$\textbf{Item}$:     Sign of GPE
$\textbf{Action}$:   Conc.Prop
$\textbf{Provided}$: Reference point and location of object
$\textbf{Outcome}$:  Correctly determine the sign of the GPE of an obejct
\end{lstlisting}
\end{minipage} \\
\midrule
\vspace{-5mm}
\begin{minipage}[t]{\linewidth}
\begin{lstlisting}[mathescape=true,basicstyle=\ttfamily\normalsize]
$\textbf{LO Code}$:  $\colorbox{lightyellow}{\textbf{ME-CME-1}}$
$\textbf{LO Name}$:  Consv. of ME (CME)
$\textbf{Item}$:     Application of Conservation of Mechanical Energy
$\textbf{Action}$:   Proc.app
$\textbf{Provided}$: A physical situation
$\textbf{Outcome}$:  Apply conservation of mechanical energy to the problem 
\end{lstlisting}
\end{minipage} \\

\bottomrule
\end{tabular}
\end{adjustbox}
\vspace{5mm}
\caption{Examples of LOs in the Energy chapter including LO Code, LO Name, Item, Action, Provided and Outcome.}
\label{tab:los_example}
\end{table}

Each Learning Objective (LO) consists of three core components: \textbf{\textit{Provided}}, \textbf{\textit{Action}}, and \textbf{\textit{Outcome}} .
The \textit{Provided} component contains a natural language description of the part of the problem context that is related to the LO. For example, a typical \textit{Provided} component could be ``velocity of an object". The \textit{Outcome} component uses natural language to describe the expected output from students, such as ``magnitude of the kinetic energy". The \textit{Action} component represents the cognitive process which students engage in to derive the \textit{Outcome} from the \textit{Provided} information. In our current construction system, the \textit{Action} component is restricted to only the following four types of actions:
\begin{enumerate}

    \item \textbf{Concept Identification (Conc. ID)}: Identify a physics concept / entity, and distinguish it from similar concepts / entities. For example, identifying kinetic energy from a description of object's motion.
    
    \item \textbf{Concept Property (Conc. Prop)}: Find or identify one of the properties associated with a given concept. For example, finding the magnitude of Kinetic Energy.
    
    \item \textbf{Procedure Application (Proc. App)}: Apply a procedure to solve a problem. Procedures can either involve the application of an equation, such as $\vec{F} = m\vec{a}$, or based on a set of pre-defined actions, such as selecting the appropriate coordinate system or reference point. 
    
    \item \textbf{Representation Mapping (Rep. Map)}: Map a visual or verbal representation to a concept or a math expression. For example: Mapping the area under the curve to velocity, mapping ``sliding down from the ramp” to a specific situation and mapping ``velocity stays constant” to a = 0. 
\end{enumerate}   

\begin{table}[t!]
\begin{adjustbox}{width=0.75\textwidth, center}
\begin{tabular}{c|ccc|ccc}
\toprule
\textbf{Chapter} & \multicolumn{3}{c|}{\textbf{Number of LO Codes}} & \multicolumn{3}{c}{\textbf{Number of LO Names}} \\
\midrule
\multirow{4}{*}{\textbf{Newton's Laws}} & \multirow{4}{*}{41} & Conc. ID & 14 & \multirow{4}{*}{16} & Physics & 8 \\
 &  & Conc. Prop & 8 &  & Representation & 1 \\
 &  & Proc. App & 17 &  & Special Case & 7 \\
 &  & Rep. Map & 2 &  &  &  \\
\midrule
\multirow{4}{*}{\textbf{Energy}} & \multirow{4}{*}{20} & Conc. ID & 5 & \multirow{4}{*}{10} & Physics & 7 \\
 &  & Conc. Prop & 5 &  & Representation & 1 \\
 &  & Proc. App & 7 &  & Special Case & 2 \\
 &  & Rep. Map & 3 &  &  &  \\
\midrule
\multirow{4}{*}{\textbf{Linear Momentum}} & \multirow{4}{*}{18} & Conc. ID & 6 & \multirow{4}{*}{6} & Physics & 3 \\
 &  & Conc. Prop & 5 &  & Representation & 0 \\
 &  & Proc. App & 7 &  & Special Case & 3 \\
 &  & Rep. Map & 0 &  &  &  \\
\bottomrule
\end{tabular}
\end{adjustbox}
\vspace{5mm}
\caption{Taxonomy of expert-curated learning objectives from the 3 selected chapters for labeling. The taxonomy for the remaining chapters can be found in Appendix A.}
\label{tab:los}
\end{table}

The distinction between \textit{Concept Property} and \textit{Procedure Application} is that \textit{Concept Property} only involves a single concept, such as Kinetic Energy in the ``Energy” chapter, and the \textit{Outcome}, such as magnitude, can be seen as a property of the concept. Procedure Application frequently involve more than one focused concept, and the final \textit{Outcome} cannot be seen as a property any of the involved concepts. 

These four types of actions are designed to be at the atomic level of cognitive process, which allows a combination of multiple atomic LOs to be mapped to higher level taxonomy systems such as the Bloom's taxonomy \cite{Gogus2012} or NTEO \cite{Teodorescu2013}. It is important to note that while these actions are the most common processes required in solving typical chapter-related problems, they do not represent an exhaustive list of the atomic level cognitive processes. 

An example of LO for kinetic energy has the components: \textit{Provided: velocity of an object, Action: Conc.Prop, Outcome: magnitude of the kinetic energy}, which can be interpreted as: \textit{Given the velocity of an object, the students should be able to find the magnitude of kinetic energy of the object, which is a property of the concept kinetic energy.} 

Each learning objective is also associated with an \textbf{\textit{Item}} property, which contains a short natural language description of the specific LO, such as ``Magnitude of KE”. Additionally, every LO is assigned a \textbf{\textit{LO Name}} property that corresponds to a general physics concept, such as ``Kinetic Energy.” Each \textit{LO name} could contain multiple learning objectives and falls into three categories: 
\begin{enumerate}
    \item \textbf{Physics Laws}: fundamental physics laws and physics concepts, such as Newton’s Second Law
    \item \textbf{Representations}: important representations used in physics, such as x-t graph or Free Body Diagram
    \item \textbf{Special Cases}: important special application of physics laws commonly covered in courses, such as inclined plane, Atwoods machine, circular motion (as a special case for Newton’s second law)
\end{enumerate}

The property of \textit{Item} and \textit{LO Name} are designed for human experts to efficiently search and organize the list of LOs. Finally, each LO is also labeled with a unique \textbf{\textit{LO Code}} (e.g., ME-KE-2), for quick reference and tagging by human experts. For instance, the code "ME-KE-2" stands for "mechanical energy topic, kinetic energy concept, second learning objective". Some detailed examples of the learning objectives in the Energy chapter are shown in Table~\ref{tab:los_example}. 

In our study, we curated 9 chapters of learning objectives, with each chapter containing 15-40 specific learning objectives. We proposed in a total of 194 distinct learning objectives (\textit{LO Codes}), spanning 76 unique general physics concepts (\textit{LO Names}). The detailed breakdown for three selected chapters can be found in Table~\ref{tab:los}, with the remaining chapters provided in Appendix A.

\subsection{Questions}
\begin{table}[t!]
\begin{adjustbox}{width=0.8\textwidth, center}
\begin{tabular}{cccc}
\toprule
\textbf{Chapter} & \textbf{Source} & \textbf{Dataset} & \textbf{Number}\\
\midrule
 Energy & Course & Energy &  9\\
 Energy & OpenStax & Chapter 8 &  44\\
\midrule
 Newton's Law & Course & Forces &  12\\
 Newton's Law & Course & Newton's 2nd Law of Motion (N2L) &  11\\
\midrule
 Linear Momentum & OpenStax & Chapter 9 &  55\\

\bottomrule
\end{tabular}
\end{adjustbox}
\vspace{5mm}
\caption{Number of selected questions for expert and LLMs labeling across chapters and sources.}
\label{tab:question}
\end{table}

We selected a set of 131 questions across three chapters: \textbf{Energy}, \textbf{Newton's Laws}, and \textbf{Linear Momentum}. The questions and chapters are selected to balance the time constraints of experts, diversity of topics, and moderate number of LOs. For the purpose of the current study, we include the questions that tests only the LOs covered in the given chapter, and avoid problems that require LOs from other chapters. The questions are selected from two main sources, an introductory college-level physics course and the end-of-chapter review questions for OpenStax\footnote{OpenStax is a worldwide publisher of open education resources: \url{https://openstax.org/}.} University Physics textbook. Table~\ref{tab:question} provides a summary of the question distribution.

\subsection{Experts Tagging Process}

The expert responsible for developing the LOs also handles the labeling LOs of all questions, which minimizes inconsistencies of the LOs labeling process. The expert labeled LOs for each questions are used as the ground truth for automated LOs labeling with LLMs.

\begin{figure}[t!]
\centering
\includegraphics[width=0.9\linewidth]{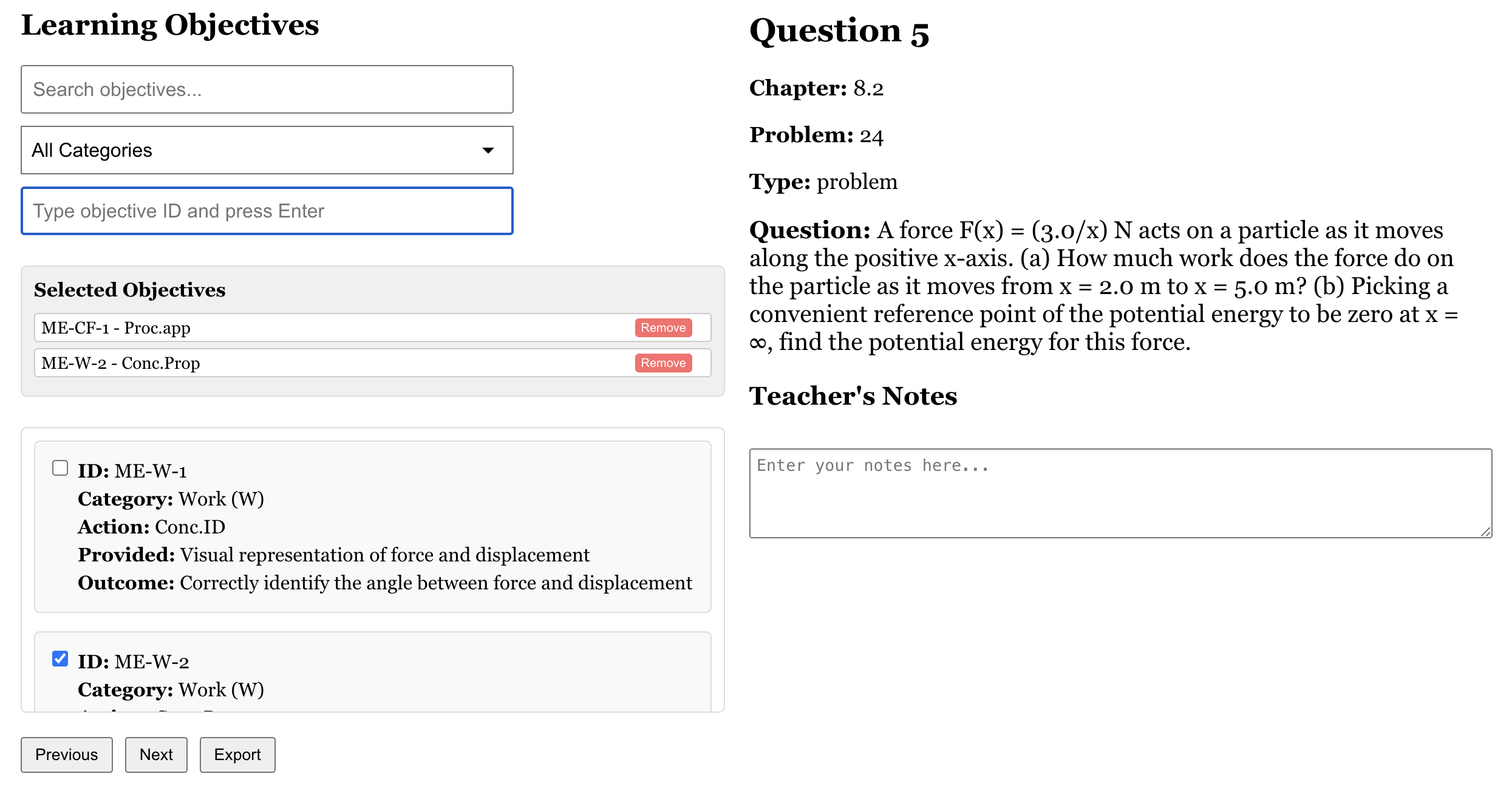} 
\vspace{2mm}
\caption{An illustration of the interface developed for human experts to label learning objectives.}
\label{fig:website}
\end{figure}

Additionally, We have developed an interface, as illustrated in Figure~\ref{fig:website} to facilitates efficient labeling by human experts. The interface displays detailed information about each question, such as the chapter and problem type, alongside a searchable database of predefined LOs. Each LO entry includes \textit{LO Codes}, \textit{LO Names} (Categories), \textit{Action}, \textit{Provided}, and \textit{Outcome}, which allows human to easily search and attach the appropriate LOs to each question. As demonstrated in Figure~\ref{fig:website}, human can select relevant LOs by either searching for specific IDs or browsing through categories. Once an LO is selected, it appears under the ``Selected Objectives” section, where it can be reviewed or removed as necessary. Additionally, the interface provides a space for the expert to enter notes for each question, which supports detailed documentation of the decision-making process.

\section{Experiments}
\label{section:exp}
In this section, we explore various LLMs on labeling atomic learning objectives. We provide our experimental setup and detailed quantitative analysis of the results.

\subsection{Experimental Setup}

\begin{table}[t]
\begin{center}
\fbox{
\begin{minipage}{\dimexpr\textwidth-1cm}
\normalsize
\texttt{
You are provided with: \\
1. A list of physics learning objectives in the following format: [INSERT FORMAT] \\
2. A physics question \\
\textcolor{blue}{
\textbf{---------------------------------------- Simple Prompting -------------------------------------} \\
Your task is to analyze the question and select ALL relevant learning objectives. \\
The instructions are: \\
1. Carefully read the physics question and the list of learning objectives. \\
2. Select ALL learning objectives that are directly applicable to solving or understanding the question. \\}
\textcolor{red}{
\textbf{--------------------------------- Explanation Prompting ---------------------------------} \\
Your task is to analyze the question and select ALL relevant learning objectives. Please provide brief explanation for each selection. \\
The instructions are: \\
1. Carefully read the physics question and the list of learning objectives. \\
2. Select ALL learning objectives that are directly applicable to solving or understanding the question \\
3. For each selected objective, provide a concise explanation of its relevance to the question. \\}
\textcolor{forestgreen}{
\textbf{------------------------------------- CoT Prompting -----------------------------------------} \\
Your task is to analyze the question and select ALL relevant learning objectives. Use chain-of-thought reasoning with step-by-step thought process. Please provide brief explanation for each selection. \\
The instructions are: \\
1. Carefully read the physics question and the list of learning objectives. \\
2. Provide step-by-step thought process to analyze the question and think about its relevance with the provided learning objectives. \\
3. Based on the thought process, select ALL learning objectives that are directly applicable to solving or understanding the question. \\
4. For each selected objective, provide a concise explanation of its relevance to the question. \\}
The list of learning objectives:
[INSERT LEARNING OBJECTIVES] \\
The physics problem: 
[INSERT THE QUESTION] \\
Important: \\
- Select ALL relevant objectives, not just the most relevant one. \\
- There are usually multiple relevant objectives for each question.
}
\end{minipage}
}
\end{center}
\vspace{2mm}
\caption{Detailed prompts for automated LOs labeling with LLMs.}
\label{tab:prompts}
\end{table}

We conduct experiments on the Learning Objective labeling task with four state-of-the-art LLMs: GPT-4o, GPT-3.5 from OpenAI \cite{achiam2023gpt} and LLaMA-70B and LLaMA-8B from Meta\footnote{The specific models we use are GPT-4o: \textit{gpt-4o-2024-05-13}, GPT-3.5: \textit{gpt-3.5-turbo-0125}, LLaMA-70B: \textit{Meta-LLaMA3.1-70B-Instruct}, LLaMA-8B: \textit{Meta-LLaMA3.1-8B-Instruct}} \cite{dubey2024llama}. For all models, we use a temperature of 0.9  to encourage diverse reasoning paths and top-p value of 1 to control robustness of the generation. We explore three different prompting strategies, with the specific prompts provided in Table~\ref{tab:prompts}. 
\begin{enumerate}
    \item \textbf{Simple prompting}: LLMs are instructed to only select the LOs without any justifications or reasoning steps.
    \item \textbf{Explanation prompting}: LLMs are instructed to select all LOs first and then then providing explanations to justify their selections. This strategy emphasizes implicit reasoning.
    \item \textbf{CoT prompting}: Building on Explanation prompting, LLMs are asked to provide a step-by-step reasoning process (chain-of-thought) before selecting any LOs, which emphasizes explicit reasoning. \end{enumerate}

Additionally, we test two types of different format to formulate learning objectives.
\begin{enumerate}
    \item \textbf{Structured LO format}, where LOs are presented systematically, as shown in the Table~\ref{tab:los_example}. The structured format follows the pattern: \textbf{[Code]: [LO Name], [Description], Provided: [Given Information], Outcome: [Expected Outcome].} 
    \item \textbf{Natural Language LO format}, where LOs are expressed more conversationally by integrating the “\textit{Provided}” and “\textit{Outcome}” elements into an action-based explanation. This format can be represented as \textbf{[Code]: [LO Name], [Description], Explanation: [provided information to a student and the expected outcome]}. For example, ME-W-1: LO Name: Work (W), Description: angle between force and distance, Explanation: Given visual representation of force and displacement, the student should be able to correctly identify the angle between force and displacement.
\end{enumerate}

We provide the evaluation results for the automated learning objective labeling task in Table~\ref{tab:result-energy}, ~\ref{tab:result-forces}, ~\ref{tab:result-lm}, corresponding to the chapter of Energy, Newton's Laws and Linear Momentum respectively. Due to the length constraints, we only include the results for GPT-4o and LLaMA-70B, with the results for GPT-3.5 and LLaMA-8B shown in Appendix B.

\subsection{Analysis of Evaluation Metrics}
Exact Matches are rare across all models, chapters, and datasets, occurring in only a small fraction of cases. This rarity highlights the challenging nature of precisely labeling every LOs with LLMs. Both the Jaccard Index and F1 score shows similar trends, but higher Jaccard and F1 scores does not always align with lower Distance values. For instance, in the ``OpenStax - ch8" dataset, LLaMA-70B achieves better Jaccard and F1 scores but yields higher Distance values compared to GPT-4o. This difference suggests that while LLaMA-70B is relatively better at capturing the exact LOs, it struggles more with understanding the hierarchical structure of LOs.

\begin{table}[t!]
\begin{adjustbox}{width=\textwidth, center}
\small
\begin{tabular}{cccccccc}
\toprule
\textbf{Dataset} & \textbf{Model} & \textbf{Prompting} & \textbf{Format} & \textbf{EM $\uparrow$} & \textbf{Jaccard $\uparrow$} & \textbf{F1 $\uparrow$} & \textbf{Distance $\downarrow$} \\
\midrule
\multirow{6}{*}{\textbf{Course - Energy}} & \multirow{6}{*}{GPT-4o} & Simple & 1 & 0 & 0.483 & 0.644
 & 3.889 \\
 &  & Explanation & 1 & $\sfrac{1}{9}$ & 0.581 & 0.708 & 3.667 \\
 &  & CoT & 1 & 0 & 0.571 & 0.714 & 3.444 \\
 &  & Simple & 2 & 0 & 0.489 & 0.646 & 5.111 \\
 &  & Explanation & 2 & 0 & 0.522 & 0.668 & 4.222 \\
 &  & CoT & 2 & $\sfrac{1}{9}$ & 0.597 & 0.732 & 3.000 \\
\midrule
 \multirow{6}{*}{\textbf{Course - Energy}} & \multirow{6}{*}{LLaMA-70B} & Simple & 1 & 0 & 0.572 & 0.720 & 4.000 \\
 &  & Explanation & 1 & 0 & 0.616 & 0.750 & 3.000 \\
 &  & CoT & 1 & 0 & 0.658 & 0.786 & 3.000 \\
 &  & Simple & 2 & 0 & 0.594 & 0.739 & 4.111 \\
 &  & Explanation & 2 & 0 & 0.625 & 0.761 & 2.778 \\
 &  & CoT & 2 & $\sfrac{3}{9}$ & 0.683 & 0.788 & 3.222 \\
\midrule
 \multirow{6}{*}{\textbf{OpenStax - ch8}} & \multirow{6}{*}{GPT-4o} & Simple & 1 & $\sfrac{2}{44}$ & 0.503 & 0.651 & 4.773 \\
 &  & Explanation & 1 & $\sfrac{1}{44}$ & 0.534 & 0.676 & 4.523 \\
 &  & CoT & 1 & $\sfrac{1}{44}$ & 0.535 & 0.676 & 4.568 \\
 &  & Simple & 2 & $\sfrac{2}{44}$ & 0.516 & 0.659 & 4.773 \\
 &  & Explanation & 2 & $\sfrac{1}{44}$ & 0.540 & 0.682 & 4.455 \\
 &  & CoT & 2 & $\sfrac{2}{44}$ & 0.559 & 0.698 & 4.432 \\
 \midrule
 \multirow{6}{*}{\textbf{OpenStax - ch8}} & \multirow{6}{*}{LLaMA-70B} & Simple & 1 & 0 & 0.509 & 0.653 & 6.068 \\
 &  & Explanation & 1 & $\sfrac{1}{44}$ & 0.551 & 0.688 & 5.045 \\
 &  & CoT & 1 & $\sfrac{2}{44}$ & 0.540 & 0.677 & 5.636 \\
 &  & Simple & 2 & $\sfrac{1}{44}$ & 0.557 & 0.696 & 5.545 \\
 &  & Explanation & 2 & $\sfrac{1}{44}$ & 0.553 & 0.697 & 5.182 \\
 &  & CoT & 2 & $\sfrac{2}{44}$ & 0.531 & 0.669 & 5.636 \\
\bottomrule
\end{tabular}
\end{adjustbox}
\vspace{5mm}
\caption{Performance results of the LO labeling task using GPT-4o and LLaMA-70B for questions in the Energy chapter.}
\label{tab:result-energy}
\end{table}
\begin{table}[t!]
\begin{adjustbox}{width=\textwidth, center}
\begin{tabular}{cccccccc}
\toprule
\textbf{Dataset} & \textbf{Model} & \textbf{Prompting} & \textbf{Format} & \textbf{EM $\uparrow$} & \textbf{Jaccard $\uparrow$} & \textbf{F1 $\uparrow$} & \textbf{Distance $\downarrow$} \\
\midrule
\multirow{6}{*}{\textbf{Course - Forces}} & \multirow{6}{*}{GPT-4o} & Simple & 1 & 0 & 0.286 & 0.414 & 10.545 \\
 &  & Explanation & 1 & 0 & 0.320 & 0.463 & 10.818 \\
 &  & CoT & 1 & 0 & 0.289 & 0.415 & 10.182 \\
 &  & Simple & 2 & 0 & 0.319 & 0.433 & 14.091 \\
 &  & Explanation & 2 & $\sfrac{1}{11}$ & 0.269 & 0.402 & 10.636 \\
 &  & CoT & 2 & 0 & 0.295 & 0.447 & 10.000 \\
\midrule
 \multirow{6}{*}{\textbf{Course - Forces}} & \multirow{6}{*}{LLaMA-70B} & Simple & 1 & $\sfrac{1}{11}$ & 0.369 & 0.493 & 9.818 \\
 &  & Explanation & 1 & 0 & 0.303 & 0.434 & 8.364 \\
 &  & CoT & 1 & 0 & 0.369 & 0.501 & 7.727 \\
 &  & Simple & 2 & 0 & 0.337 & 0.486 & 8.545 \\
 &  & Explanation & 2 & 0 & 0.281 & 0.429 & 8.636 \\
 &  & CoT & 2 & 0 & 0.345 & 0.490 & 8.000 \\
\midrule
 \multirow{6}{*}{\textbf{Course - N2L}} & \multirow{6}{*}{GPT-4o} & Simple & 1 & 0 & 0.247 & 0.378 & 10.333 \\
 &  & Explanation & 1 & 0 & 0.225 & 0.337 & 10.750 \\
 &  & CoT & 1 & 0 & 0.311 & 0.448 & 7.250 \\
 &  & Simple & 2 & 0 & 0.217 & 0.323 & 10.250 \\
 &  & Explanation & 2 & 0 & 0.213 & 0.336 & 10.917 \\
 &  & CoT & 2 & 0 & 0.291 & 0.435 & 9.167 \\
 \midrule
 \multirow{6}{*}{\textbf{Course - N2L}} & \multirow{6}{*}{LLaMA-70B} & Simple & 1 & 0 & 0.267 & 0.384 & 9.917 \\
 &  & Explanation & 1 & 0 & 0.326 & 0.474 & 8.000 \\
 &  & CoT & 1 & 0 & 0.311 & 0.417 & 8.083 \\
 &  & Simple & 2 & 0 & 0.225 & 0.350 & 10.583 \\
 &  & Explanation & 2 & 0 & 0.267 & 0.395 & 8.667 \\
 &  & CoT & 2 & 0 & 0.321 & 0.471 & 8.250 \\
\bottomrule
\end{tabular}
\end{adjustbox}
\vspace{2mm}
\caption{Performance results of the LO labeling task using GPT-4o and LLaMA-70B for questions in the Newton's Laws chapter.}
\label{tab:result-forces}
\end{table}
\begin{table}[t!]
\begin{adjustbox}{width=\textwidth, center}
\begin{tabular}{cccccccc}
\toprule
\textbf{Dataset} & \textbf{Model} & \textbf{Prompting} & \textbf{Format} & \textbf{EM $\uparrow$} & \textbf{Jaccard $\uparrow$} & \textbf{F1 $\uparrow$} & \textbf{Distance $\downarrow$} \\
\midrule
\multirow{6}{*}{\textbf{OpenStax - ch9}} & \multirow{6}{*}{GPT-4o} & Simple & 1 & $\sfrac{2}{55}$ & 0.432 & 0.584 & 4.800 \\
 &  & Explanation & 1 & 0 & 0.412 & 0.564 & 4.855 \\
 &  & CoT & 1 & $\sfrac{1}{55}$ & 0.440 & 0.597 & 4.782 \\
 &  & Simple & 2 & $\sfrac{1}{55}$ & 0.444 & 0.593 & 4.764 \\
 &  & Explanation & 2 & $\sfrac{1}{55}$ & 0.478 & 0.626 & 4.364 \\
 &  & CoT & 2 & $\sfrac{1}{55}$ & 0.468 & 0.617 & 4.309 \\
\midrule
 \multirow{6}{*}{\textbf{OpenStax - ch9}} & \multirow{6}{*}{LLaMA-70B} & Simple & 1 & $\sfrac{1}{55}$ & 0.320 & 0.457 & 5.691 \\
 &  & Explanation & 1 & 0 & 0.376 & 0.516 & 5.400 \\
 &  & CoT & 1 & 0 & 0.354 & 0.502 & 5.382 \\
 &  & Simple & 2 & $\sfrac{1}{55}$ & 0.357 & 0.500 & 5.891 \\
 &  & Explanation & 2 & 0 & 0.356 & 0.499 & 5.364 \\
 &  & CoT & 2 & 0 & 0.338 & 0.477 & 5.491 \\
\bottomrule
\end{tabular}
\end{adjustbox}
\vspace{2mm}
\caption{Performance results of the LO labeling task using GPT-4o and LLaMA-70B for questions in the Linear Momentum chapter.}
\label{tab:result-lm}
\end{table}

\subsection{Comparison of LLMs}
Our results indicate that LLaMA-70B demonstrates comparable performance to GPT-4o, with some variations across datasets. LLaMA-70B shows better performance on the course dataset, while GPT-4o has a slight advantage in OpenStax dataset. This difference may be attributed to inherent structural differences between two datasets. Both GPT-4o and LLaMA-70B show satisfactory performance (with F1 value of 0.5-0.7) for for the Energy and Linear Momentum chapters, indicating the potential of LLMs in automate LO labeling. However, the Newton's Laws chapter still remains challenging for all models (with F1 value of 0.3-0.4), potentially due to the higher number of LOs in this chapter (41 compared to 20 LOs in Energy) or the increased complexity of mathematical reasoning required in solving questions related to Newton's Laws. Additionally, our results reveal that smaller models (results shown in Appendix B), such as GPT-3.5 and LLaMA-8B occasionally outperform larger models, particularly when evaluated with the distance metric. This phenomenon is especially evident in the Newton's Laws section, where larger models tend to over-select LOs, leading to higher distances. This tendency may be a result of larger language model's broader knowledge base, which, while generally advantageous, can lead to reduced precision under specific contexts.

\subsection{Effectiveness of Prompting and Formatting Strategies}
Our results demonstrate that Chain-of-Thought (CoT) prompting generally outperforms Explanation prompting, which in turn outperforms Simple prompting. This trend is more consistent for GPT-4o than LLaMA-70B, possibly due to GPT-4o's superior structured reasoning ability. However, Explanation prompting sometimes performs worse than Simple prompting, particularly in Newton's Laws chapter (e.g. approximately 0.06 decrease for \textit{Course-Forces} with LLaMA-70B), while in Linear Momentum chapter, CoT occasionally underperforms Explanation prompting (e.g. approximately 0.02 decrease for \textit{Openstax-ch9} with LLaMA-70B). Notably, Simple prompting rarely outperforms CoT directly, occurring in only 1/20 instances in Table~\ref{tab:result-energy}, ~\ref{tab:result-forces}, ~\ref{tab:result-lm}. The results underscore that both implicit (Explanation) and explicit (CoT) reasoning can enhance model performance, and their combination often yields the best results.

Additionally, our results reveal that both structured and natural language LO formatting achieve comparable overall performance, with domain-specific variations. For Energy and Linear Momentum chapters, natural language formatting generally yield better performance, whereas for Newton's Laws, structured formatting is more effective. The difference is possibly due to the nature of the concepts: Energy and Linear Momentum benefit from descriptive language, whereas the rule-based concepts in Newton's Laws are better suited to a structured format. Interestingly, natural language formatting appear to enhance Simple prompting the most, likely because it incorporates more contextual information. However, this additional information seems less important for more sophisticated prompting strategies, which already encourage deeper reasoning processes.

\section{Discussions}
In this section, we provide analysis and visualization for the number of LOs per question, the prediction frequency and accuracy for each LO. We also conduct a qualitative error analysis for LLMs labeling results. 

\begin{table}[t!]
\begin{tabular}{c|c|ccc|ccc|ccc|ccc}
\toprule
\textbf{} & \multirow{2}{*}{\textbf{Human}} & \multicolumn{3}{c|}{\textbf{GPT-4o}} & \multicolumn{3}{c|}{\textbf{GPT-3.5}} & \multicolumn{3}{c|}{\textbf{LLaMA-70B}} & \multicolumn{3}{c}{\textbf{LLaMA-8B}} \\
\textbf{} &  & \textbf{Sim} & \textbf{Exp} & \textbf{CoT} & \textbf{Sim} & \textbf{Exp} & \textbf{CoT} & \textbf{Sim} & \textbf{Exp} & \textbf{CoT} & \textbf{Sim} & \textbf{Exp} & \textbf{CoT} \\
\midrule
\textbf{Energy} & 5.27 & 4.91 & 5.07 & 4.93 & 2.55 & 2.55 & 3.24 & 6.80 & 6.27 & 6.35 & 5.53 & 5.89 & 5.15 \\
\textbf{Newton's Laws} & 3.48 & 8.13 & 7.22 & 6.70 & 3.13 & 2.61 & 3.48 & 6.83 & 6.57 & 6.26 & 6.17 & 5.70 & 6.00 \\
\textbf{Linear Momentum} & 4.47 & 4.18 & 4.76 & 4.40 & 1.89 & 2.11 & 2.71 & 5.71 & 5.02 & 4.65 & 3.67 & 4.24 & 4.33 \\
\bottomrule
\end{tabular}
\vspace{2mm}
\caption{
Average number of learning objectives labeled per question by humans and various LLMs across all chapters.}
\label{tab:number_lo}
\end{table}
\begin{figure*}[t!]
\centering
\includegraphics[width=0.8\linewidth]{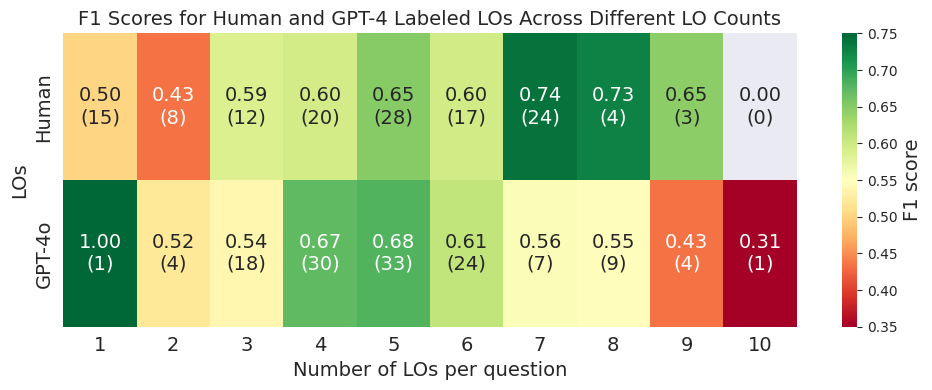} 
\vspace{2mm}
\caption{An illustration of F1 score across various number of LOs per question for human and GPT-4o labeled LOs. The top number represents F1 score and the bottom number represents question counts. For instance, the upper-left box indicates 15 questions are labeled by human experts with only 1 LO and those 15 questions achieve an average F1 score of 0.5. The prompting strategy adopted is CoT prompting with natural language LO format (best viewed in colors).}
\label{fig:heatmap}
\end{figure*}

\subsection{Number of Learning Objectives per Question}
As shown in Table~\ref{tab:number_lo}, each question is associated with approximately 4 LOs on average. Interestingly, the chapter of Newton's Laws has the least average LOs per question (3.48) despite having the highest total number of LOs (41). This counter-intuitive result may explain why LLMs tend to over-predict LOs for questions in this chapter, resulting in reduced performance. Larger models generally predict more LOs compared to smaller models, which has both benefits and drawbacks. While it may lead to more comprehensive concept coverage, particularly for complex questions, it also risks including irrelevant objectives that could mislead students. Additionally, LLaMA models consistently predict more LOs than comparably sized GPT models.

The heatmap in Figure~\ref{fig:heatmap} reveals that human labeled questions primarily have 4-7 LOs, while GPT-4o typically assigns 4-6 LOs per question, indicating a reasonable alignment in LO distribution. However, F1 scores are lower for questions with very few (1-3) or many (9-10) LOs, implying that models struggle with extreme cases. Both human and GPT-4o labels achieve the highest F1 scores in the middle range (5-7 LOs), where most questions fall. This pattern highlights the challenge of accurately labeling questions with LO counts that deviate significantly from the average.

\begin{figure*}[t!]
\begin{minipage}[t]{0.49\linewidth}
\begin{center}
\includegraphics[width=1\linewidth]{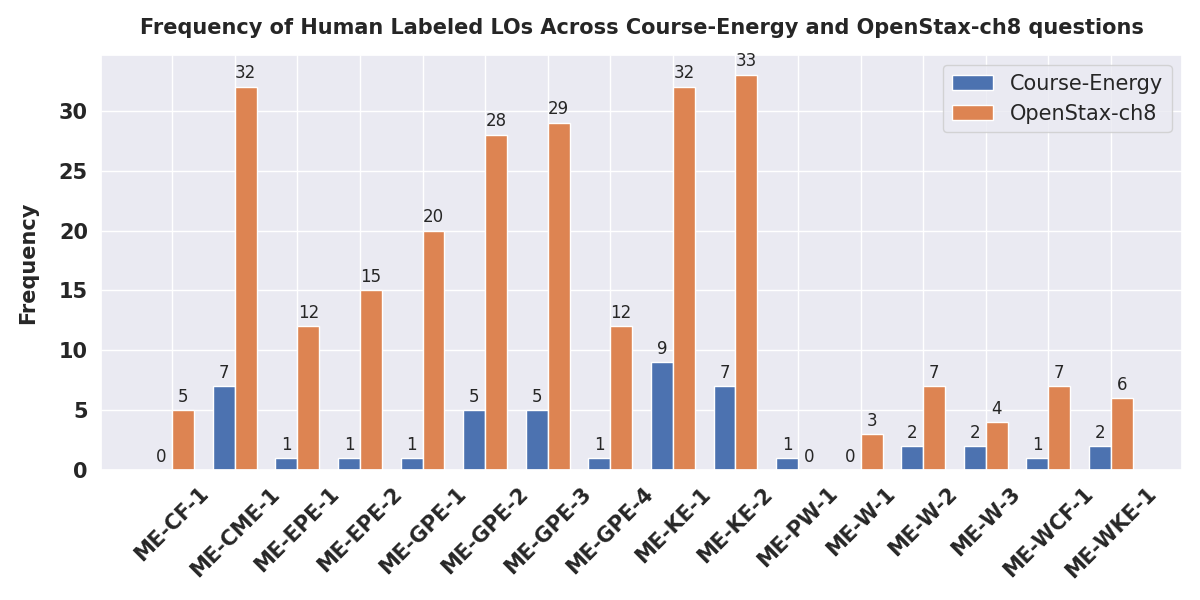} 
\end{center} 
\end{minipage}
\hfill
\hspace{-5mm}
\begin{minipage}[t]{0.49\linewidth}
\begin{center}
\includegraphics[width=1\linewidth]{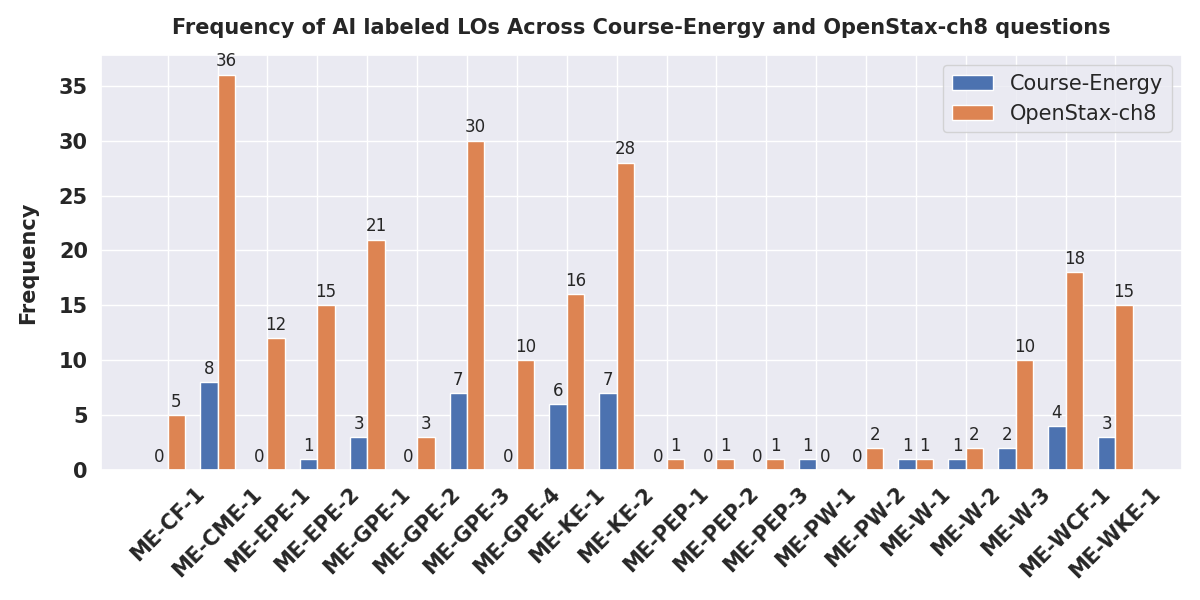} 
\end{center}
\end{minipage}
\vspace{2mm}
\caption{Frequency distribution of LOs in the Energy chapter: human-labeled LOs on the left, GPT-4o generated LOs with CoT prompting and natural language format on the right.}
\label{fig:frequency}
\end{figure*}
\begin{figure*}[t!]
\centering
\includegraphics[width=1\linewidth]{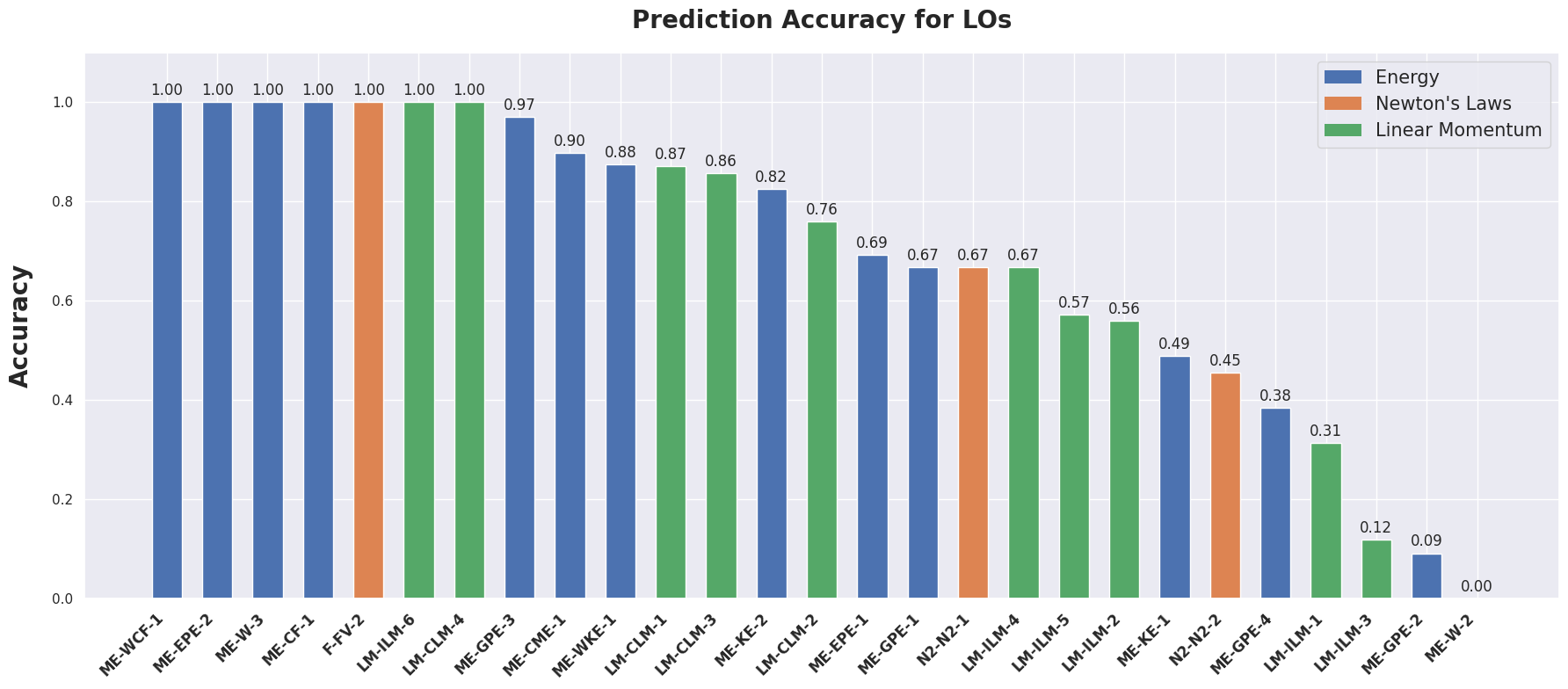} 
\vspace{2mm}
\caption{Accuracy of LOs labeled by GPT-4 using CoT prompting and natural language LO format. Only LOs appearing in five or more questions are included.}
\label{fig:accuracy}
\end{figure*}

\subsection{Frequency and Accuracy per Learning Objective}

Figure~\ref{fig:frequency} compares the frequency distribution of LOs for the Energy chapter between human and GPT-4o labels. Both of them show similar distribution trends, indicating a consistent coverage between course materials and OpenStax textbook. However, GPT-4o occasionally identifies LOs that human experts did not use at all, such as the ME-PEP series, which pertains to a special physics representation called the ``potential energy profile" graph.

Figure~\ref{fig:accuracy} further illustrates the GPT-4o's accuracy in labeling LOs that appear in at least five questions. The model's performance varies significantly across LOs, with seven achieving 1.0 accuracy, while one is completely missed. Of the 7 LOs that had accuracy of 1.0, the action component of 6 of those is procedure application. This shows that GPT-4o is highly accurate in determining the core physics principle or the main equation that needs to be applied to common textbook questions. On the other hand, of the five LOs with the lowest arrucacy, four of which involve action of either concept identification or concept property. Moreover, three of the five LOs involve determining either the direction of a quantity, or the sign of a vector quantity (which is essentially determining the direction). The one procedure application LO (ME-GPE-4) was not the application of a physics principle, but rather the application of a heuristic procedure of selecting a proper reference height for gravitational potential energy, which is also highly dependent on the geometric relation of the problem. This suggests that even as a multi-modal model, GPT-4o has difficulty determining features related to spatial or geometric properties of the problem.

\subsection{Error Analysis for LLMs-labeled Learning Objectives}

We conducted qualitative analysis with a human expert to evaluate GPT-4o's performance and reasoning in labeling LOs. The assessment was based on the LOs labeled by GPT-4o, the explanation for each LO and thought process. Several examples are presented below, with misclassified LOs highlighted in \textcolor{red}{red} and \textcolor{blue}{blue}.
\begin{enumerate}

    \item \texttt{\textbf{Question}: A skater of mass 40 kg is carrying a box of mass 5 kg. The skater has a speed of 5 m/s with respect to the floor and is gliding without any friction on a smooth surface. Find the momentum of the box with respect to the floor. Find the momentum of the box with respect to the floor after she puts the box down on the frictionless skating surface.} \\
    \texttt{\textbf{Human labeled LOs}: LM-ILM-2}	\\
    \texttt{\textbf{LLMs-labeled LOs}: LM-ILM-2, \textcolor{blue}{LM-CLM-1}, \textcolor{blue}{LM-CLM-2}, \textcolor{blue}{LM-CLM-4}, \textcolor{blue}{LM-CLM-5}, \textcolor{blue}{LM-ILM-6}} \\
    \texttt{\textbf{Expert Analysis}: The problem text is superficially similar to problems that involve conservation of linear momentum, but actually only involves calculating the magnitude of the linear momentum. The AI tagged with multiple Los related to conservation of linear momentum, even adding a tag related to impulse-momentum theory. However, at least some of the tags related to conservation of linear momentum might be justifiable. Technically speaking, the problem could be solved by writing down the conservation of linear momentum equation for the system, and then realizing that the equation is trivial and unnecessary. It is plausible that some students actually do so.}
    
    \item \texttt{\textbf{Question}: A mouse of mass 200 g falls 100 m down a vertical mine shaft and lands at the bottom with a speed of 8.0 m/s. During its fall, how much work is done on the mouse by air resistance?} \\
    \texttt{\textbf{Human labeled LOs}: ME-KE-1, ME-KE-2, ME-WCF-1, \textcolor{red}{ME-W-2}, \textcolor{red}{ME-W-1}, \textcolor{red}{ME-WKE-1}}	\\
    \texttt{\textbf{LLMs-labeled LOs}: ME-KE-1, ME-KE-2, \textcolor{blue}{ME-GPE-1}, \textcolor{blue}{ME-GPE-3}, ME-WCF-1, \textcolor{blue}{ME-CME-1}} \\
    \texttt{\textbf{Expert Analysis}: Human expert tagged the problem with Los related to Work (ME-W-1) and Kinetic Energy Theorem (ME-WKE-1), whereas AI tagged the problem with Conservation of Mechanical Energy (ME-CME-1), and LOs related to change in gravitational potential energy (ME-GPE-1, ME-GPE-3). Upon review, the human expert acknowledged overlooking the change in gravitational potential energy, and miss-classified the LOs. The AI tagging is more accurate.}

    \item \texttt{\textbf{Question}: A motorcycle of weight w = 2300 N is traveling at a speed of v = 24.0 m/s, following a straight line. The brakes are applied at a speed limit sign, which is equivalent to a net force of F = 3000 N against the motorcycle's displacement over a distance of d = 11.4 m. Find the motorcycle's final speed, i.e. at the end of that 11.4 m stretch.} \\
    \texttt{\textbf{Human labeled LOs}: ME-WKE-1, \textcolor{blue}{ME-W-2}, ME-W-3, ME-KE-2, ME-KE-1}	\\
    \texttt{\textbf{LLMs-labeled LOs}: ME-W-3, ME-WKE-1, ME-KE-1, ME-KE-2} \\
    \texttt{\textbf{Expert Analysis}: AI correctly identified all the concepts involved. It missed ME-W-2, which is related to determining the sign of the work done. The likely reason is that the LO explicitly states ``Provided the angle between force and displacement", and the problem didn't explicitly provide an angle, but implicitly inferred an 180 degree angle by stating that the brakes are being applied.}
\end{enumerate}

We hypothesize four main reasons that could lead to incorrect labeling by LLMs. First, the current LO system is developed for human experts to use, and therefore the description sometimes ``cuts corners", and rely on the experience and implicit knowledge of human experts. This compression of information, while reducing cognitive load for human experts, can lead to confusion for AI systems, such as LOs related to ``conservation of mechanical energy" where similar procedures are condensed into a single LO.

Second, in some cases, human experts often use implicit rules when labeling LOs. For example, human experts could label LOs related to concept identification when the concept is not explicitly mentioned in the problem. These implicit rules are not effectively conveyed to the LLMs through the description of LOs.

Third, LLMs frequently struggle with LOs pertaining to spatial relation of objects in the problem. For instance, LOs related to determining the height for gravitational potential energy is often miss-labeled, as the decision depends on the understanding of the geometrical relation within the problem.

Forth, in some cases, LLMs have a tendency to omit LOs related to implicit steps in the solution, such as missing LOs related to the direction and magnitude of linear momentum in linear momentum conservation problems. However, this pattern of omission is not frequently observed for Energy-related problems, where LLMs consistently tags LOs related to kinetic energy of an object, even though the two situations are largely parallel. This type of error could potentially be reduced by either providing LLMs with the solution, or asking LLMs to generate a solution, and subsequently generate LO labels based on both the problem and the solution.

On the other hand, the rare cases in which LLMs tagged LOs that the human expert initially overlooked demonstrates the significant advantage of AI: its immunity to fatigue and carelessness. This is especially critical for labeling atomic LOs, as the process is far more mentally challenging for humans compared to labeling problems with conventional LOs. 

\section{Conclusion and Future Directions}
This paper introduces a novel atomic learning objective system designed to capture the detailed cognitive processes and concepts essential for physics problem-solving. Our extensive experiments and evaluations demonstrate the potential of LLMs in categorizing physics problems with high resolution. We observed that possible reasons for the LLMs mislabeling could be attributed to the LO system's original design, which is intended for human users rather than AI models. 

Future research directions include expanding the atomic LO system to cover a wide range of chapters and subjects. We also aim to continuously improve our atomic LO system to enable AI to carry out most of the labeling, with human experts focusing on quality assurance and system refinement. Additionally, we want to explore whether including problem solutions could potentially improve LLMs' labeling accuracy. Finally, we aim to explore LLMs' capability to generate questions based on selected learning objectives.

\section*{Acknowledgments}
This work was supported by NSF grant 1842378, ONR grant N0014-20-1-2534, AFOSR grant FA9550-22-1-0060, and a Vannevar Bush Faculty Fellowship, ONR grant N00014-18-1-2047 and MURI N00014-20-1-2787. The authors would like to thank UCF Department of Physics for supporting the development of the initial set of learning objectives.

\bibliographystyle{abbrv}  
\bibliography{reference}  

\newpage
\appendix
\section{Taxonomy of learning objectives}
We curated 9 chapters of learning objectives, with each chapter containing 15-40 specific learning objectives. The detailed taxonomy for all chapters can be found in Table~\ref{tab:los_all}.

\begin{table}[h]
\begin{adjustbox}{width=0.95\textwidth, center}
\begin{tabular}{c|ccc|ccc}
\toprule
\textbf{Chapter} & \multicolumn{3}{c|}{\textbf{Number of LO Codes}} & \multicolumn{3}{c}{\textbf{Number of LO Names}} \\
\midrule
\multirow{4}{*}{1D Motion} & \multirow{4}{*}{15} & Conc. ID & 3 & \multirow{4}{*}{6} & Physics & 4 \\
 &  & Conc. Prop & 6 &  & Representation & 2 \\
 &  & Proc. App & 1 &  & Special Case & 0 \\
 &  & Rep. Map & 5 &  &  &  \\
\midrule
\multirow{4}{*}{2D Motion} & \multirow{4}{*}{21} & Conc. ID & 3 & \multirow{4}{*}{8} & Physics & 6 \\
 &  & Conc. Prop & 7 &  & Representation & 0 \\
 &  & Proc. App & 11 &  & Special Case & 2 \\
 &  & Rep. Map & 0 &  &  &  \\
\midrule
\multirow{4}{*}{\textbf{Newton's Laws}} & \multirow{4}{*}{41} & Conc. ID & 14 & \multirow{4}{*}{16} & Physics & 8 \\
 &  & Conc. Prop & 8 &  & Representation & 1 \\
 &  & Proc. App & 17 &  & Special Case & 7 \\
 &  & Rep. Map & 2 &  &  &  \\
\midrule
\multirow{4}{*}{\textbf{Energy}} & \multirow{4}{*}{20} & Conc. ID & 5 & \multirow{4}{*}{10} & Physics & 7 \\
 &  & Conc. Prop & 5 &  & Representation & 1 \\
 &  & Proc. App & 7 &  & Special Case & 2 \\
 &  & Rep. Map & 3 &  &  &  \\
\midrule
\multirow{4}{*}{\textbf{Linear Momentum}} & \multirow{4}{*}{18} & Conc. ID & 6 & \multirow{4}{*}{6} & Physics & 3 \\
 &  & Conc. Prop & 5 &  & Representation & 0 \\
 &  & Proc. App & 7 &  & Special Case & 3 \\
 &  & Rep. Map & 0 &  &  &  \\
\midrule
\multirow{4}{*}{Rotational Dynamics} & \multirow{4}{*}{28} & Conc. ID & 8 & \multirow{4}{*}{8} & Physics & 5 \\
 &  & Conc. Prop & 7 &  & Representation & 0 \\
 &  & Proc. App & 13 &  & Special Case & 3 \\
 &  & Rep. Map & 0 &  &  &  \\
\midrule
\multirow{4}{*}{Angular Momentum and Angular Collision} & \multirow{4}{*}{14} & Conc. ID & 3 & \multirow{4}{*}{5} & Physics & 3 \\
 &  & Conc. Prop & 4 &  & Representation & 0 \\
 &  & Proc. App & 7 &  & Special Case & 2 \\
 &  & Rep. Map & 0 &  &  &  \\
\midrule
\multirow{4}{*}{Simple Harmonic Oscillation} & \multirow{4}{*}{17} & Conc. ID & 3 & \multirow{4}{*}{7} & Physics & 2 \\
 &  & Conc. Prop & 1 &  & Representation & 2 \\
 &  & Proc. App & 7 &  & Special Case & 3 \\
 &  & Rep. Map & 6 &  &  &  \\
\midrule
\multirow{4}{*}{Waves and Universal Law of Gravity} & \multirow{4}{*}{20} & Conc. ID & 4 & \multirow{4}{*}{10} & Physics & 6 \\
 &  & Conc. Prop & 3 &  & Representation & 1 \\
 &  & Proc. App & 10 &  & Special Case & 3 \\
 &  & Rep. Map & 3 &  &  &  \\
\bottomrule
\end{tabular}
\end{adjustbox}
\vspace{2mm}
\caption{Taxonomy of expert-curated learning objectives from 9 chapters. Bolded chapters include questions selected for LOs labeling.}
\label{tab:los_all}
\end{table}

\newpage
\section{Additional Quantitative results}
We provide the evaluation results for the automated learning objective labeling task with smaller models GPT-3.5 and LLaMA-8B in Table~\ref{tab:result-energy-small},~\ref{tab:result-forces-small},~\ref{tab:result-lm-small}, corresponding to the chapter of Energy, Newton’s Laws and Linear Momentum respectively. The analysis for the results is shown in Section~\ref{section:exp}.

\begin{table}[h]
\begin{adjustbox}{width=\textwidth, center}
\begin{tabular}{cccccccc}
\toprule
\textbf{Dataset} & \textbf{Model} & \textbf{Prompting} & \textbf{Format} & \textbf{EM $\uparrow$} & \textbf{Jaccard $\uparrow$} & \textbf{F1 $\uparrow$} & \textbf{Distance $\downarrow$} \\
\midrule
\multirow{6}{*}{\textbf{Course - Energy}} & \multirow{6}{*}{GPT-3.5} & Simple & 1 & 0 & 0.202 & 0.309 & 7.667 \\
 &  & Explanation & 1 & 0 & 0.245 & 0.367 & 7.000 \\
 &  & CoT & 1 & 0 & 0.333 & 0.476 & 5.889 \\
 &  & Simple & 2 & 0 & 0.248 & 0.365 & 7.444 \\
 &  & Explanation & 2 & 0 & 0.257 & 0.399 & 6.778 \\
 &  & CoT & 2 & 0 & 0.325 & 0.474 & 5.778 \\
\midrule
 \multirow{6}{*}{\textbf{Course - Energy}} & \multirow{6}{*}{LLaMA-8B} & Simple & 1 & 0 & 0.399 & 0.537 & 7.333 \\
 &  & Explanation & 1 & 0 & 0.333 & 0.485 & 6.778 \\
 &  & CoT & 1 & 0 & 0.409 & 0.569 & 6.000 \\
 &  & Simple & 2 & 0 & 0.350 & 0.492 & 7.889 \\
 &  & Explanation & 2 & 0 & 0.388 & 0.540 & 7.444 \\
 &  & CoT & 2 & 0 & 0.449 & 0.605 & 6.000 \\
\midrule
 \multirow{6}{*}{\textbf{OpenStax - ch8}} & \multirow{6}{*}{GPT-3.5} & Simple & 1 & 0 & 0.247 & 0.365 & 8.000 \\
 &  & Explanation & 1 & 0 & 0.213 & 0.322 & 8.500 \\
 &  & CoT & 1 & 0 & 0.283 & 0.419 & 7.545 \\
 &  & Simple & 2 & 0 & 0.243 & 0.345 & 8.091 \\
 &  & Explanation & 2 & 0 & 0.239 & 0.352 & 8.114 \\
 &  & CoT & 2 & $\sfrac{1}{44}$ & 0.286 & 0.411 & 7.795 \\
 \midrule
 \multirow{6}{*}{\textbf{OpenStax - ch8}} & \multirow{6}{*}{LLaMA-8B} & Simple & 1 & $\sfrac{1}{44}$ & 0.332 & 0.463 & 8.750 \\
 &  & Explanation & 1 & 0 & 0.351 & 0.488 & 8.114 \\
 &  & CoT & 1 & 0 & 0.424 & 0.557 & 6.955 \\
 &  & Simple & 2 & 0 & 0.344 & 0.483 & 8.773 \\
 &  & Explanation & 2 & 0 & 0.370 & 0.511 & 8.250 \\
 &  & CoT & 2 & 0 & 0.397 & 0.534 & 7.455 \\
\bottomrule
\end{tabular}
\end{adjustbox}
\vspace{2mm}
\caption{Performance results of the LO labeling task using GPT-3.5 and LLaMA-8B for questions in the Energy chapter.}
\label{tab:result-energy-small}
\end{table}
\begin{table}[t!]
\begin{adjustbox}{width=\textwidth, center}
\begin{tabular}{cccccccc}
\toprule
\textbf{Dataset} & \textbf{Model} & \textbf{Prompting} & \textbf{Format} & \textbf{EM $\uparrow$} & \textbf{Jaccard $\uparrow$} & \textbf{F1 $\uparrow$} & \textbf{Distance $\downarrow$} \\
\midrule
\multirow{6}{*}{\textbf{Course - Forces}} & \multirow{6}{*}{GPT-3.5} & Simple & 1 & 0 & 0.205 & 0.306 & 5.727 \\
 &  & Explanation & 1 & $\sfrac{1}{11}$ & 0.222 & 0.287 & 6.273 \\
 &  & CoT & 1 & 0 & 0.183 & 0.285 & 6.636 \\
 &  & Simple & 2 & 0 & 0.159 & 0.249 & 7.545 \\
 &  & Explanation & 2 & 0 & 0.145 & 0.221 & 6.545 \\
 &  & CoT & 2 & 0 & 0.233 & 0.339 & 7.000 \\
\midrule
 \multirow{6}{*}{\textbf{Course - Forces}} & \multirow{6}{*}{LLaMA-8B} & Simple & 1 & 0 & 0.226 & 0.321 & 10.000 \\
 &  & Explanation & 1 & 0 & 0.234 & 0.354 & 9.727 \\
 &  & CoT & 1 & 0 & 0.193 & 0.299 & 9.364 \\
 &  & Simple & 2 & 0 & 0.220 & 0.328 & 10.364 \\
 &  & Explanation & 2 & 0 & 0.246 & 0.363 & 8.636 \\
 &  & CoT & 2 & 0 & 0.246 & 0.358 & 9.091 \\
\midrule
 \multirow{6}{*}{\textbf{Course - N2L}} & \multirow{6}{*}{GPT-3.5} & Simple & 1 & 0 & 0.165 & 0.252 & 6.167 \\
 &  & Explanation & 1 & 0 & 0.196 & 0.294 & 6.417 \\
 &  & CoT & 1 & 0 & 0.167 & 0.260 & 6.667 \\
 &  & Simple & 2 & 0 & 0.147 & 0.221 & 6.833 \\
 &  & Explanation & 2 & 0 & 0.151 & 0.209 & 7.250 \\
 &  & CoT & 2 & 0 & 0.183 & 0.259 & 7.500 \\
 \midrule
 \multirow{6}{*}{\textbf{Course - N2L}} & \multirow{6}{*}{LLaMA-8B} & Simple & 1 & 0 & 0.150 & 0.247 & 10.417 \\
 &  & Explanation & 1 & 0 & 0.180 & 0.271 & 11.167 \\
 &  & CoT & 1 & 0 & 0.204 & 0.294 & 10.250 \\
 &  & Simple & 2 & 0 & 0.170 & 0.265 & 10.750 \\
 &  & Explanation & 2 & 0 & 0.205 & 0.306 & 9.667 \\
 &  & CoT & 2 & 0 & 0.151 & 0.238 & 10.583 \\
\bottomrule
\end{tabular}
\end{adjustbox}
\vspace{2mm}
\caption{Performance results of the LO labeling task using GPT-3.5 and LLaMA-8B for questions in the Newton's Laws chapter.}
\label{tab:result-forces-small}
\end{table}
\begin{table}[t!]
\begin{adjustbox}{width=\textwidth, center}
\begin{tabular}{cccccccc}
\toprule
\textbf{Dataset} & \textbf{Model} & \textbf{Prompting} & \textbf{Format} & \textbf{EM $\uparrow$} & \textbf{Jaccard $\uparrow$} & \textbf{F1 $\uparrow$} & \textbf{Distance $\downarrow$} \\
\midrule
\multirow{6}{*}{\textbf{OpenStax - ch9}} & \multirow{6}{*}{GPT-3.5} & Simple & 1 & 0 & 0.130 & 0.191 & 7.855 \\
 &  & Explanation & 1 & $\sfrac{2}{55}$ & 0.203 & 0.291 & 7.382 \\
 &  & CoT & 1 & $\sfrac{1}{55}$ & 0.272 & 0.391 & 5.818 \\
 &  & Simple & 2 & $\sfrac{1}{55}$ & 0.125 & 0.182 & 7.709 \\
 &  & Explanation & 2 & 0 & 0.152 & 0.232 & 7.673 \\
 &  & CoT & 2 & 0 & 0.216 & 0.336 & 6.255 \\
\midrule
 \multirow{6}{*}{\textbf{OpenStax - ch9}} & \multirow{6}{*}{LLaMA-8B} & Simple & 1 & $\sfrac{1}{55}$ & 0.219 & 0.323 & 6.564 \\
 &  & Explanation & 1 & 0 & 0.235 & 0.352 & 7.000 \\
 &  & CoT & 1 & $\sfrac{1}{55}$ & 0.236 & 0.346 & 6.818 \\
 &  & Simple & 2 & $\sfrac{2}{55}$ & 0.228 & 0.332 & 6.745 \\
 &  & Explanation & 2 & 0 & 0.213 & 0.321 & 7.655 \\
 &  & CoT & 2 & 0 & 0.242 & 0.353 & 6.618 \\
\bottomrule
\end{tabular}
\end{adjustbox}
\vspace{2mm}
\caption{Performance results of the LO labeling task using GPT-3.5 and LLaMA-8B for questions in the Linear Momentum chapter.}
\label{tab:result-lm-small}
\end{table}

\end{document}